\documentclass[prl,twocolumn,superscriptaddress]{revtex4}

\usepackage{graphicx}
\usepackage{dcolumn}
\usepackage{bm}
\usepackage{url}
\newcommand{\be}{\begin{equation}}
\newcommand{\ee}{\end{equation}}
\newcommand{\ba}{\begin{eqnarray}}
\newcommand{\ea}{\end{eqnarray}}
\newcommand{\nn}{\nonumber \\}

\newcommand{\bra}[1]{\left\langle{#1}\right\vert}
\newcommand{\ket}[1]{\left\vert{#1}\right\rangle}
\begin{document}

\preprint{DWAVE/TAQC-01}
\title{Thermally assisted adiabatic quantum computation}
\author{M.~H.~S.~Amin}
\email{amin@dwavesys.com}
\affiliation{D-Wave Systems Inc., 100-4401 Still Creek Drive, Burnaby, B.C., V5C 6G9, Canada}
\author{Peter~J.~Love}
\affiliation{D-Wave Systems Inc., 100-4401 Still Creek Drive,
Burnaby, B.C., V5C 6G9, Canada} \affiliation{Department of Physics,
Haverford College, 370 Lancaster Avenue, Haverford, PA 19041-1392,
USA}
\author{C.~J.~S.~Truncik}
\affiliation{D-Wave Systems Inc., 100-4401 Still Creek Drive, Burnaby, B.C., V5C 6G9, Canada}


\begin{abstract}

We study the effect of a thermal environment on adiabatic quantum
computation using the Bloch-Redfield formalism. We show that in
certain cases the environment can enhance the performance in two
different ways: (i) by introducing a time scale for thermal mixing
near the anticrossing that is smaller than the adiabatic time scale,
and (ii) by relaxation after the anticrossing. The former can
enhance the scaling of computation when the environment is
superohmic, while the latter can only provide a prefactor
enhancement. We apply our method to the case of adiabatic Grover
search and show that performance better than classical is possible
with a superohmic environment, with no {\it a priori} knowledge of
the energy spectrum.

\end{abstract}

\maketitle


Quantum computation (QC) aims to harness the physical resources made
available by quantum mechanics to gain an advantage over classical
computation. A major obstacle to construction of a large scale
quantum computer is loss of coherence resulting from uncontrolled
coupling to the environment. In principle, environmental effects may
be circumvented by the use of quantum error
correction~\cite{Nielsen,
bib:steaneqecc,bib:shorft,bib:gottesmanft}. In practice, however,
such schemes require significant overhead. It is therefore likely
that many noisy qubits will be available before many error-corrected
logical qubits are.

This observation motivates the search for models of QC with
intrinsic resistance to noise. One such example is adiabatic quantum
computation (AQC)~\cite{Farhi,aharonov2004,Childs}. Here we
investigate a regime in which weak coupling to an environment can
improve the performance of AQC.

In AQC, information is stored in the ground state of a quantum
system and manipulated by control of the system Hamiltonian. An AQC
is operated by deforming an initial Hamiltonian $H_i$ into a final
Hamiltonian $H_f$ through intermediates $ H_S = [1-\lambda(t)]H_i +
\lambda(t)H_f$, with $\lambda(t)$ changing from 0 to 1 between the
initial ($t_i{=}0$) and final ($t_f$) times. If the evolution
satisfies the adiabatic condition ($\hbar{=}k_B{=}1$ throughout):
$|\langle 1|dH/dt|0\rangle| \ll g^2(\lambda)$, where $g(\lambda)$ is
the energy gap between the ground ($\ket{0}$) and first excited
($\ket{1}$) states, then the system will be in the ground state of
$H_f$ at $t_f$ with probability close to one, and the solution may
then be read out \footnote{A generalized adiabatic condition for
open quantum systems was introduced by M.~S.~Sarandy and
D.~A.~Lidar, Phys. Rev. A {\bf 71}, 012331 (2005); Phys. Rev. Lett.
{\bf 95}, 250503 (2005).}. In a global adiabatic scheme,
$\lambda=t/t_f$ and the adiabatic condition must be satisfied for
the smallest gap $g_m$. If $g(\lambda)$ is known, one can choose
$d\lambda/dt \propto g^2(\lambda)$ to enhance the performance using
a local adiabatic scheme \cite{Roland}. Here, we assume no {\it a
priori} knowledge of the energy spectrum, and use $\lambda=t/t_f$
throughout. The amount of time required to successfully run a
computation is determined by the minimum gap between the first two
energy levels, $g_m$, along the path connecting $H_i$ and $H_f$. In
order for the evolution to remain adiabatic throughout, the total
time required is $t_f \propto 1/g_m^2$.

Here we analyze the behavior of an AQC in the presence of a thermal
environment with temperature $T \gg g_m$. We restrict our analysis
to problems in which the performance is limited by a single minimum
gap of the type of an energy level avoided crossing. This
corresponds to a first order quantum phase transition, which is
believed to be hardest for AQC \cite{Schutzhold06}.

In general, if there are $l$ energy levels within the range $T$ from
the ground state, then thermalization can suppress the ground state
probability by at most a factor of $l^{-1}$. For a Gaussian
distribution of the levels, $l$ is polynomial in the number of
qubits $n$, if $T$ is much smaller than the total spectral width. In
this case, one may compensate for thermalization by repetition  with
a polynomial overhead. Moreover, the transition times are expected
to be very long, probably longer than the computation time,
otherwise classical annealing would yield the solution efficiently.
This is different from an anticrossing for which, as we shall see,
the transition rate is sharply peaked at the anticrossing point. We
therefore only focus on the anticrossing and use 2-level
approximation.

Let us assume that the minimum gap occurs at $\lambda{=}\lambda_m$.
We adopt a new coordinate, $\epsilon {=} 2E(\lambda{-}\lambda_m) $,
where $E$ is an energy scale characterizing the anticrossing. Close
to the anticrossing, the system Hamiltonian within the 2-level
approximation is well described by
\be H_S =-(\epsilon \tau_z + g_m \tau_x )/2 ,
 \label{H2L}
\ee
and the gap between the first two states is well approximated by $g
= \sqrt{\epsilon^2 + g_m^2}$. Here $\tau_{x,z}$ are the Pauli
matrices in the 2-level subspace. Due to the Landau-Zener transition
\cite{Landau32,Zener32}, the probability of being in the excited
state at $t=t_f$ is given by
 \be
 P_{1f} = e^{-t_f/t_a}, \label{closed}
 \ee
where $t_a=4E/\pi g_m^2$ is the adiabatic time scale (see Table
\ref{table} for definition of all time scales).

We incorporate the environment by assuming that qubits are coupled
to bosonic heat baths that are in equilibrium with $g_m \ll T \ll
E$. The total Hamiltonian is $H=H_S+H_B+H_{\rm int}$, where $H_B$
and $H_{\rm int}$ are bath and interaction Hamiltonians
respectively. We also assume that in the 2-level subspace the
interaction Hamiltonian has the form
 \be
 H_{\rm int} = Q \otimes \tau_z, \label{Hint}
 \ee
where $Q$ is an operator representing the collective effect of all
baths on the 2-state problem.
Equations (\ref{H2L}) and (\ref{Hint}) capture the physics of a wide range of problems that have one sharp anticrossing.

\begin{table}
\caption{\label{table} Characteristic time scales for $t_f$}
\begin{ruledtabular}
\begin{tabular}{ll}
$t_a$ & time scale imposed by non-adiabatic transitions \\
$t_d$ & time scale imposed by thermal mixing \\
$t_r$ & time scale imposed by relaxation after the anticrossing \\
\end{tabular}
\end{ruledtabular}
\end{table}

For slow evolutions of the Hamiltonian considered here, as long as
the correlation time of the environment is shorter than decay times
of the system, one can safely assume Markovian approximation
\footnote{As noted by R. Alicki, D.A. Lidar, and P. Zanardi,
Phys.~Rev.~A {\bf 73}, 052311 (2006), Markovian approximation
requires the relaxation rates to be much smaller than the minimum
gap, hence $\Gamma(0) \ll g_m$. Using (\ref{rhozdot}), we find
$\tilde{\gamma}(0)\gg g_m$, consistent with our original
assumptions.}.
Writing the density matrix as $\rho=(1 + \bm{\rho}
\cdot \bm{\tau})/2$, the 2-state Bloch-Redfield equations are
\cite{Blum}:
\ba
 \dot \rho_x &=& - \tilde{\gamma} \rho_x + \epsilon \rho_y
 - \left({\epsilon \over g_m} \gamma - {g_m \over \epsilon} \gamma_\varphi \right)  \rho_z
 + \gamma {g \over g_m} \rho_{\rm eq}, \nn
 \dot\rho_y &=& -\epsilon \rho_x - \tilde{\gamma} \rho_y + g_m \rho_z, \label{eqns} \\
 \dot\rho_z &=&  - g_m \rho_y, \nonumber
\ea
where $\gamma = (g_m/g)^2 [S(g) + S(-g)]$, $\gamma_\varphi = 2
(\epsilon/g)^2 S(0)$, $\tilde{\gamma}=\gamma + \gamma_\varphi$, and
$\rho_{\rm eq} \equiv [S(g)-S(-g)]/[S(g) + S(-g)]$. Here, the bath's
spectral density is defined as $S(\omega)=\int_{-\infty}^\infty dt \
e^{i\omega t} \langle Q(t) Q(0) \rangle$, where $\langle ...
\rangle$ denotes averaging over environmental degrees of freedom.
The prefactor $(g_m/g)^2$ makes $\gamma$ sharply peaked at
$\epsilon=0$ as expected.

For a bosonic environment \cite{Leggett}, $S(\omega) =
J(\omega)/(1-e^{-\omega/T})$, where $J(\omega) = \eta \omega
\left|{\omega/\omega_c}\right|^s e^{-\omega/\omega_c}$,
with $\omega_c$ being a cutoff frequency which is assumed to be
larger than all other relevant energy scales in the system.
Therefore
\ba
 \gamma = (g_m/g)^2 J(g) \coth (g/2T), \quad
 \rho_{\rm eq} = \tanh (g/2T).  \label{gamma}
\ea
Here, we focus only on ohmic ($s=0$) and superohmic ($s>0$) cases
for which the correlation time of the bath ($\sim 1/\omega_c$) is
short compared to the relevant time scales. A subohmic ($s<0$)
environment has a large correlation time, hence the Markovian
approximation and therefore Bloch-Redfield equation do not hold
\cite{Tiersch}.

\begin{figure}[t]
\includegraphics[width=5cm]{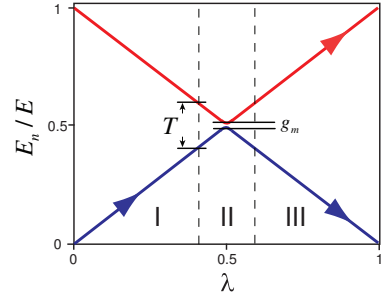}
\caption{Ground and first excited states of a system with one
anticrossing at $\lambda=0.5$. The 3 regions are identified in such
a way that $g>T$ for regions I and III, and $g<T$ for region II.}
\label{fig1}
\end{figure}

We are interested in problems with small gap, $g_m {\ll} T {\ll} E$.
We divide the evolution into three regions, as shown in
Fig.~\ref{fig1}. In region I, the gap is larger than $T$ and thermal
transitions are suppressed. In region II, both thermal and
non-adiabatic transitions between the two states are possible. In
region III, the system again has a gap larger than $T$, but now the
system can relax from the excited state to the ground state. Such
relaxation can only increase the probability of success.

Let us start by finding the excitation probability immediately after
region II. Assuming $T {\gg} g$, which holds for most of the region,
we have $\rho_{\rm eq} = 0$, and
\ba
 \gamma = \gamma_0 {g_m^2 \over g^2} \left({g \over
 \omega_c}\right)^s,
 \qquad
 \gamma_\varphi = \left\{ \begin{array}{cc}
 \gamma_0 {\epsilon^2 / g^2} & \ s=0 \\
 0 & \ s>0 \end{array} \right. \label{gammahighT}
\ea
where $\gamma_0 {=} 2\eta T$. We perform the calculation in the
regime $g_m {\ll} \tilde{\gamma}$. The presence of the damping terms
in the first two equations in (\ref{eqns}) will make $\rho_x$ and
$\rho_y$ decay in a time scale ($\sim 1/\tilde{\gamma}$) much
shorter than the relevant time scale for $\rho_z$ ($\sim 1/g_m$).
Thus, to find the slow evolution of $\rho_z$, one can use the
stationary values for $\rho_x$ and $\rho_y$, obtained from
$\dot\rho_{x,y} = 0$. Solving the first two equations in
(\ref{eqns}) for $\rho_y$ and substituting into the third equation,
we get
 \be
 \dot\rho_z = - \Gamma(\epsilon) \rho_z, \qquad
 \Gamma(\epsilon)= g^2 \gamma /(\tilde{\gamma}^2 +
 \epsilon^2).  \label{rhozdot}
 \ee
Here, $\Gamma$ is the rate of transition between the two branches of
energy that meet at the anticrossing. Integrating (\ref{rhozdot}),
we find
 \ba
 \ln{\rho_z(t) \over \rho_z(0)} =
 -\int_0^{t} dt' \Gamma(\epsilon) = -
 \int_{-T}^{\epsilon(t)} {d\epsilon \over \dot\epsilon}
 \Gamma(\epsilon), \label{lnrho} 
 \ea
which leads to
 \ba
 \rho(\epsilon \approx T) = \rho(0) e^{-t_f/t_d},
 \ea
where $t_d$ is a characteristic time scale that describes the
thermal mixing near the anticrossing, and is given by
 \ba
 {1 \over t_d} =
 {1 \over t_f} \int_{-\infty}^{\infty} {d\epsilon \over \dot
 \epsilon} \Gamma(\epsilon) .
 \ea
We have taken the integration limits to infinity assuming that
$\Gamma(\epsilon)$ is sharply peaked at $\epsilon = 0$, which is the
case for $s<1$. Using the initial condition $\rho_z(0)=1$, the
excitation probability after region II is approximately
 \be
 P_{1}(\epsilon \approx T) = {1\over 2}(1+e^{-t_f/t_d}). \label{P1}
 \ee

For a linear time evolution (i.e., $\lambda=t/t_f$), one has
$\dot\epsilon = 2E/t_f$. If the environment is ohmic, then $s=0$ and
$\tilde{\gamma}=\gamma_0$ is a constant. Therefore
 \ba
 t_d \approx \left( {g_m^2 \over 2E} \int_{-\infty}^\infty
 {\gamma_0 d\epsilon  \over \gamma_0^2 + \epsilon^2}\right)^{-1} =
 {2E \over \pi g_m^2} = {t_a \over 2},
 \ea
in agreement with Kayanuma \cite{Kayanuma} and Ao and Rammer
\cite{Ao}. For fast evolutions (short $t_f$), (\ref{P1}) behaves the
same way as a closed system (\ref{closed}), while in the slow regime
(long $t_f$), (\ref{P1}) gives a ground state probability of $\sim
1/2$, corresponding to the complete mixture of the two states.

For a superohmic bath ($s>0$), $\gamma_\varphi = 0$, hence
$\tilde{\gamma} = \gamma = \gamma_0 (g_m/g)^2 (g/ \omega_c)^s$. In
such a case,
 \ba
 {1 \over t_d} = {1 \over 2E}
 \int_{-\infty}^\infty d\epsilon {\Lambda g^s \over \Lambda^2g^{2s-4} +
 \epsilon^2},
 \ea
where $\Lambda = \gamma_0 g_m^2/\omega_c^s$. The important
contribution to the integral comes from regions with $g\approx
|\epsilon| \sim \Lambda^{1/(3-s)} \gg g_m$, where the inequality
follows from $g_m \ll \gamma_0 (g_m / \omega_c)^s =
\tilde{\gamma}(0)$, which was our initial assumption. This
condition, however, can be satisfied in the limit of $g_m\to 0$,
only if $s<1$. Replacing $\epsilon = \Lambda^{1/(3-s)} x$, we find
 \be
 t_d = \alpha_s E \left( {\omega_c^s \over \gamma_0 g_m^2}
 \right)^{2\over 3-s}, \quad
 {1 \over \alpha_s} \approx
 \int_0^{\infty} {x^{4-s}dx \over 1+x^{6-2s}}.
 \ee
Since $\alpha_s$ is independent of $g_m$, we have $t_d\propto
g_m^{-4/(3-s)}$ which scales better than $t_a\sim g_m^{-2}$. It is
easy to check that the integral is convergent for $s<1$. For $s >
1$, the condition $g_m \ll \tilde{\gamma}$ cannot be satisfied in
the limit of $g_m\to 0$, invalidating our approach.

We now study the effect of relaxation after the anticrossing (region
III). From (\ref{rhozdot}), we see that $\Gamma (\epsilon {\gg}
\tilde{\gamma}) \approx \gamma$. The probability of ending up in the
excited state becomes
$
 P_{1f}(\epsilon \approx E) \approx P_1(\epsilon \approx T) e^{-\int_T^{E} \gamma
 d\epsilon/\dot\epsilon}.
$
Using (\ref{gamma}) and assuming $\coth (g/2T) \approx 1$, which
holds for most of the region, and $\epsilon \gg g_m$, we find
 \be
 {1\over 2E}\int_T^{E} \gamma d\epsilon \approx {\eta g_m^2\over 2E \omega_c^s}
 \int_T^{E} \epsilon^{s-1} d\epsilon \equiv {1\over t_r}.
 \ee
Here we have defined a third time scale $t_r$ that characterizes
such a relaxation process. One can write $t_r{=}t_a/\kappa_s$, where
 \be
 \kappa_s = {2\eta \over \pi} \left\{ \begin{array}{cc}
  \ln (E/T) & \ s=0 \\
 {1\over s}[(E/\omega_c)^s - (T/\omega_c)^s] & \ s>0
 \end{array} \right. . \label{kappa}
 \ee
Notice that $t_r$ slowly decreases with $T$.

The probability of success, i.e., the final ground state
probability, is therefore given by
 \be
 P_{0f}(t) \approx 1- {1\over 2}\left(1+e^{-t_f/t_d}\right)
 e^{-t_f/t_r}. \label{P0f}
 \ee
It reaches $\sim 1/2$ in a time $t_f\sim t_d$, but approaches 1 in a
time $t_f \sim t_r$. If $t_d < t_r$, it is advantageous to run the
system faster but repeat the process. The relevant time scale for
computation will then be $\sim 2t_d$, which for an ohmic environment
is $\sim t_a \propto 1/g_m^2$, the time scale for a closed AQC. On
the other hand, for superohmic cases with $0{<}s{<}1$, one has $t_d
\propto g_m^{-4/(3-s)}$, which shows an improved performance
compared to the closed AQC, as $g_m \to 0$. The performance becomes
better as $s$ gets closer to zero, until $s=0$ (i.e., ohmic) at
which point the low frequency part of the noise spectrum becomes
nonzero and the performance goes back to $1/g_m^2$.

This sudden change at $s=0$ is related to the sharp jump in
$S(\omega)\propto |\omega|^s$ at $\omega=0$, from a nonzero value at
$s=0$ to zero at $s>0$. However, the $S(0)$ that appears in the
definition of $\gamma_\varphi$ is not exactly zero frequency, but
really the low frequency component of the noise, i.e.,
$S({\sim}1/t_f)$. As $s$ becomes smaller, the low frequency
component gets larger and eventually dominates the $\tilde{\gamma}$
in (\ref{lnrho}), resulting in a smooth transition to the ohmic
behavior. Without the $S(0)$ term, an ohmic environment would yield
a $t_d\sim g_m^{-4/3}$ behavior. Here, a competition between pure
relaxation, which tends to enhance the performance, and pure
dephasing (due to the low frequency noise) which works against it,
is noticeable. Taking both processes into account, in the case of
ohmic environment, the performance of the system will be the same as
that for a fully coherent AQC.

For systems with $t_d>t_r$, the computation time scale will be
determined by $t_r$ and the ground state probability, for small
$t_f$, will basically have the form $P_{0f}(t) \approx 1-
e^{-t_f/t_r}$. If $t_r < t_a$, then we will again have a better
performance compared to a closed AQC. However, as we saw before,
$t_r$ has the same $g_m^{-2}$ dependence as $t_a$. Thus, any speedup
over AQC by this process can only be via a prefactor $\kappa_s$  (if
it is larger than 1). The enhancement reported in Ref.~\cite{Childs}
falls in this category since the number of qubits considered was not
large enough to obtain small $g_m$ and therefore thermal mixture at
the anticrossing.

We should emphasize that equation (\ref{kappa}) is calculated
assuming that the 2-state approximation holds for the entire range.
While this can be the case for some Hamiltonians, such as adiabatic
Grover search \cite{Roland}, it is not true in general. In fact, it
is very difficult to calculate $t_r$ for a general problem. However,
one would not expect this type of relaxation, which is equivalent to
classical annealing, give any scaling benefit over classical
computation.

We now apply our approach to the adiabatic implementation of
Grover's search algorithm~\cite{Grover,Roland}. In this case, the
explicit dependence of $g_m$ on the problem size may be obtained,
and hence all quantities may be calculated in terms of the size of
the unstructured search problem $N=2^n$. Following Roland and Cerf
\cite{Roland}, we use the Hamiltonian
$ 
 H_S=E\left[{\bf 1} -
 (1-\lambda)\ket{+}\bra{+}-\lambda\ket{m}\bra{m}\right],
$ 
where $|m\rangle$ is the marked state to be found and
$\ket{+}{=}(N)^{-1/2}\sum_{l}\ket{l}$. Defining
$\epsilon{=}E(2\lambda -1)$, the gap is $g(\epsilon) {=} \sqrt{E^2/N
+ (1-1/N)\epsilon^2}$. The minimum gap, $g_m {=} E/\sqrt{N}$, lies
at $\epsilon=0$. The third energy level, $E_2{=}E$, has ($N-2$)-fold
degeneracy. A global adiabatic algorithm ($\lambda{=}t/t_f$) results
in $t_f {\sim} N/E$~\cite{bib:adiabatic1}. Using a local adiabatic
algorithm~\cite{Roland}, one can achieve $t_f {\sim} \sqrt{N}/E$.
Because of the large degeneracy of $E_2$, the 2-level approximation
will only be valid in the temperature regime $T {\ll} E/\log N$.

\begin{figure}[t]
\includegraphics[width=0.3\textwidth]{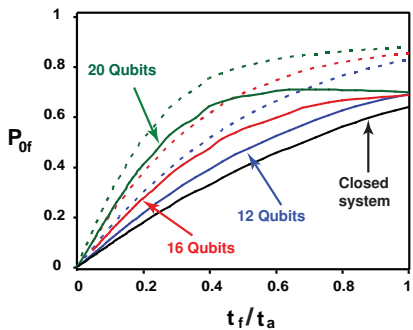}
\caption{Probability of success for AGS with 12 (blue), 16 (red),
and 20 (green) qubits, and a superohmic environment with $s{=}0.5$
at $T/E {=}0.1$. The solid and dotted curves are for longitudinal
($\bar{\eta}^x{=}0,\ \bar{\eta}^z{=}0.1$) and transverse
($\bar{\eta}^x{=}0.1,\ \bar{\eta}^z{=}0$) couplings to the
environment, respectively. The solid black line represents a closed
system ($\bar{\eta}^x {=} \bar{\eta}^z {=}0$), which with the
normalized x-axis is independent of $n$. } \label{fig2}
\end{figure}

We consider the implementation of the unstructured search problem on
n qubits, and hence may describe our noise model in terms of
operators acting on these qubits. We show that the type of 2-level
noise model (3) indeed arises for a general coupling of qubits to
the environment: $H_{\rm int} = - \sum_{i=1}^{n}
[X_i\otimes\sigma_i^x+Z_i \otimes\sigma_i^z]$, where
$\sigma_i^\alpha$ are the Pauli matrices for the $i$-th qubit, and
$X_i$, $Z_i$ are its corresponding heat bath operators. In the large
$N$ limit, the effective 2-level system and interaction Hamiltonians
become (\ref{H2L}) and (\ref{Hint}), respectively, where $Q={1 \over
2}\sum_i (X_i - Z_i)$. Assuming uncorrelated heat baths, (\ref{P0f})
also holds for this problem with $\eta = {1\over 4}n(\bar{\eta}^x +
\bar{\eta}^z)$, where $\bar{\eta}^{x,z}$ are average friction
coefficients for the $X_i$ and $Z_i$ operators.

For large $n$, the scaling of $t_d$ with $N$ is given by $t_d \sim
N$ for ohmic, and $t_d \sim N^{2/(3-s)}$ for superohmic environment
(with linear interpolation). It is clear that for superohmic
environment with $s<1$, the scaling is better than that for
classical computation.

We have also performed numerical simulations of adiabatic Grover
search, solving the Bloch-Redfield equations without the 2-level or
large $N$ approximation for $12$, $16$ and $20$ qubits.
Figure~\ref{fig2} plots $P_{0f}$ as a function of $t_f/t_a$ for a
case with superohmic environment with $s=0.5$. As is clear from the
figure, the curves increase faster compared to a closed system for
larger $n$ (smaller $g_m$). This agrees with the scaling advantage
of the noisy system compared to the closed system according to our
analytical prediction.

To summarize, using the Bloch-Redfield formalism we have identified
3 time scales for the evolution of AQC and determined their scalings
with $g_m$. We have shown that relaxation after the anticrossing can
only provide a prefactor enhancement for computation time. Thermal
mixing at the anticrossing, on the other hand, can enhance the
scaling of the computation if the environment is superohmic with
$0{<}s{<}1$, while the same environment will be destructive for gate
model QC. This underlines the important difference between the two
models in response to the environment. Finally, we should mention
that a presence of low frequency noise, as in  spin environment
\cite{Andy}, will remove the above enhancement.

\acknowledgements{Discussions with A.J.~Berkley, J.B.~Biamonte,
A.~Blais, E.~Farhi, E.~Ladizinsky, A.J.~Leggett, D.~Lidar, A.
Maassen van den Brink, D.A.~Meyer, G. Rose, A.Yu. Smirnov,
P.C.E.~Stamp, and M.~Wubs are gratefully acknowledged. MHSA would
also like to thank D.V. Averin for pointing out the importance of
low frequency noise and the breakdown of the secular approximation
in the Bloch-Redfield formalism. }

\bibliography{paper}

\end{document}